\documentclass[english,aps,prl,superscriptaddress,floatfix,notitlepage,reprint,show pacs]{revtex4-2}

\usepackage[T1]{fontenc}
\usepackage[utf8]{inputenc}
\usepackage{physics}
\usepackage{natbib}
\usepackage{amsthm}
\usepackage{float}
\usepackage{amssymb}
\usepackage{dsfont}
\usepackage{amsmath}
\usepackage{bm}
\usepackage{textcomp}
\usepackage{sublabel}
\usepackage{latexsym}
\usepackage{sidecap}
\usepackage{placeins}
\usepackage{url}
\makeatletter
\theoremstyle{plain}

\theoremstyle{plain}

\theoremstyle{plain}
\newtheorem*{prop*}{\protect\propositionname}

\usepackage{braket}
\usepackage{txfonts}
\usepackage{pifont}
\usepackage{graphicx}
\usepackage[usenames,dvipsnames]{xcolor}
\usepackage{hyperref}
\usepackage{cleveref}
\hypersetup{
    colorlinks=true,
    linkcolor=Red,       
    citecolor=blue,      
    urlcolor=cyan      
}
\usepackage{orcidlink}
\usepackage{tikz}
\usetikzlibrary{patterns,decorations.text,decorations.pathreplacing,decorations.pathmorphing}
\usepackage{caption}
\usepackage{subcaption}
\usepackage{rotating}
\usepackage{lipsum}
\date{\today}
\newcommand{\red}{\protect\tikz[baseline=-0.5ex]\draw[red] (0,0)--(0.5,0);}
\newcommand{\blue}{\protect\tikz[baseline=-0.5ex]\draw[blue] (0,0)--(0.5,0);}

\newcommand{\bluedashed}{\protect\tikz[baseline=-0.5ex]\draw[blue,dashed] (0,0)--(0.5,0);}

\begin{document}
\title{Roles of Polarization and Detuning in the Noise-induced Relaxation Dynamics of Atomic-Molecular Bose Condensates}
\author{Avinaba Mukherjee \orcidlink{0009-0000-3765-6466}}
\thanks{\href{mailto:avinaba.mukherjee@rediffmail.com}{avinaba.mukherjee@rediffmail.com}}
\author{Raka Dasgupta \orcidlink{0000-0003-2148-4641}}
\thanks{\href{mailto:rdphy@caluniv.ac.in}{rdphy@caluniv.ac.in}}
\address{Department of Physics, University of Calcutta, $92$ A. P. C. Road, Kolkata $700009$, India}
\begin{abstract}
We study the relaxation dynamics of a resonant Bose gas driven by Gaussian white noise. The system is characterized by the population imbalance between  molecules, and atoms as well as their mutual coherence. We analyze longitudinal and transverse relaxation times using mean-field theory and a Bogoliubov–Born–Green–Kirkwood–Yvon  hierarchy that incorporates higher-order correlations. We find that when the initial imbalance is near the system equilibrium, that enhances the longitudinal relaxation time, as reflected by a minimum in the drift speed near equilibrium. With an increasing value of initial polarization, the transverse relaxation time is suppressed, evidenced by a reduction in the von Neumann entropy due to coherence loss. As the Feshbach detuning is varied, the longitudinal relaxation time attains a minimum near resonance, where atom–molecule conversion is maximal. In contrast, the transverse relaxation time reaches a maximum near resonance, corresponding to maximal condensate coherence.
 \end{abstract}
\maketitle
\section{Introduction}

A coupled system of atomic and molecular Bose-Einstein condensates (BECs) has emerged as a fascinating platform, drawing significant attention and hosting various exciting quantum-statistical phenomena, in the last few decades \cite{drummond1998coherent,timmermans1999rarified,javanainen1999coherent,van1999time,yurovsky1999atom,heinzen2000superchemistry,BEC2,BEC9,BEC10,BEC11,kokkelmans2002ramsey,mackie2002mean,BEC16,BEC8,duine2003microscopic,BEC5,duine2004atom,drummond2004coherent,BEC6,BEC15,BEC7,similar_hamiltonian2,BEC1,BEC3,BEC13,BEC12,zhang2023many}. In this hybrid system, two atomic bosons may merge to form a molecular boson through a Feshbach resonance mechanism \cite{timmermans1999feshbach,kohler2006production}. This setup effectively mirrors the behavior of a Bosonic Josephson junction \cite{milburn1997quantum,BJJ1,BJJ7,BJJ8,BJJ9,BJJ18,BJJ20}, where the atom-molecule coupling serves an analogous role to the tunneling in traditional Josephson junctions. Here, the atomic (A-BEC) and molecular (M-BEC) condensate states are energetically separated by a formation threshold similar to the insulating barrier in a typical Josephson junction. This energy threshold becomes a crucial, tunable parameter that governs that the system's behaviour.\\

The study of non-equilibrium dynamics for two-mode condensates has gained significant relevance, especially in the context of condensates in a double-well geometry \cite{ref22,ref28,SBJJ}. One particular class of such out-of-equilibrium systems is a two-mode BEC undergoing relaxation dynamics \cite{stochastic_bosonic_josephson_junction,BJJ1}. The relaxation dynamics in ultracold systems can be directly probed in
experiments by measuring density distributions and currents \cite{trotzky2012probing}. The relaxation dynamics of a quantum system is extremely informative. It not only probes how the system approaches a steady state, but also reveals key information about its intrinsic properties, including kinetic constraints \cite{mori2018thermalization,darkwah2022probing} which in
turn allows one to investigate whether and how the system
thermalizes \cite{mori2018thermalization}.\\ 

In the presence of dissipation, the dynamics of a quantum system can be analyzed using a Lindblad master equation in the Markovian limit. If the system–bath coupling is weak, it can alternatively be modelled as Gaussian white noise \cite{dutta2025introduction}.
 In a double-well BEC, this noise affects the amplitude of hopping and detuning, and the Bosonic Josephson Junction exhibits damped oscillations of ``polarization", i.e.,  population imbalance, and phase coherence,  relaxing toward a steady state \cite{stochastic_bosonic_josephson_junction}. Incorporating quantum fluctuations beyond the Mean Field (MF) approach using the Bogoliubov backreaction (BBR) or truncated Bogoliubov–Born–Green–Kirkwood–Yvon (BBGKY)  formalism can successfully describe how particle coherence deteriorates due to many-body effects \cite{Linblad_master_equation}. Although the MF approximation adequately captures short-time relaxation, accommodating longer time-scales especially under phase noise necessitates the inclusion of higher order correlations in the form of the BBR method.  The same can be attempted in atomic and molecular BECs, as that is also another prominent example of
a bimodal BEC. In fact, for the combined atom-molecule system, the underlying nonlinearity has a more interesting form, resulting in richer dynamics.\\

In this article, we investigate how the initial population imbalance  and detuning affect the relaxation times associated with both population imbalance and coherence. To interpret these findings, we analyze several physical quantities, and relate them with the relaxation dynamics. For example, the effect of a varying initial population imbalance has been connected with drift speed and Von Neumann entropy; while indicators such as  conversion-efficiency and Condensate fraction have been studied to highlight the influence of a variable detuning.\\

This paper is organized as follows. Sec. \ref{framework} presents the basic two-state model and the inclusion of noise. Sec. \ref{Relaxation dynamics thermodynamic quantum} reports on the relaxation processes within both the MF and BBR descriptions. Sec. \ref{energy or particle transfer} is devoted to setting up experimentally realizable parameters. In Sec. \ref{long-trans}, we analyze the dependence of the longitudinal and transverse relaxation times on the initial polarization, and the Feshbach detuning. In Sec. \ref{intution}, several macroscopic and statistical observables characterizing the condensate state are studied, and their connections with the relaxation process is established.  We  summarize our conclusions in Sec. \ref{conclusion}. 

\section{Basic Hamiltonian Framework and Noise Effects}\label{framework}

We study a system in which two bosonic atoms can coherently bind to form a bosonic molecule through the mechanism of Feshbach resonance \cite{AMBEC2,AMBEC3}. The underlying physics is captured by a two-channel model, comprising an open (or entrance) channel and a closed channel \cite{AMBEC3}. When the energy level of the open channel aligns with that of the closed channel, a resonant coupling occurs, allowing two free atoms to form a bound molecular state i.e., a bosonic dimer. The energy gap between the A-BEC and the M-BEC is denoted by $\epsilon_b$. This resonance condition can be precisely controlled by varying an external magnetic field, effectively tuning the relative energies of the two channels. The coupled atom–molecule dynamics are governed by the following toy Hamiltonian:
\cite{similar_hamiltonian3,similar_hamiltonian4}.\\
\begin{equation}
\label{hamiltonian}
    \begin{split}
        \hat{H}= & \frac{u_1}{2V} \hat{a}^\dagger \hat{a}^\dagger \hat{a}\hat{a}+ \frac{u_2}{2V} \hat{b}^\dagger \hat{b}^\dagger \hat{b}\hat{b}+ \frac{u_3}{V} \hat{a}^\dagger \hat{b}^\dagger \hat{b}\hat{a}\\ &+ \frac{g}{\sqrt{V}}(\hat{a}^\dagger \hat{a}^\dagger \hat{b}+ \hat{b}^\dagger \hat{a}\hat{a}) + \epsilon_b  \hat{b}^\dagger \hat{b}
    \end{split}
\end{equation}
Here $\hat{a}^\dagger$ and $\hat{a}$ denote the creation and annihilation operators for bosonic atoms, respectively, while $\hat{b}^\dagger$ and $\hat{b}$ serve as the corresponding operators for bosonic molecules. The parameters $u_1$ and $u_2$ characterize the interaction strengths between atoms and between molecules, respectively. The term $u_3$ accounts for the interaction between atomic and molecular bosons. The coupling constant $g$ characterizes the strength of the Feshbach resonance responsible for the conversion between atomic and molecular states, where $V$ denotes the quantization volume. Here, both the A-BEC and M-BEC are treated as single-mode condensates. Eq. (\ref{hamiltonian}) serves as a toy model that captures the essential features of harmonically trapped BECs in the weak-interaction limit, neglecting external trap-induced spatial inhomogeneities and the thermal population of excited states. This, in spirit, is similar to using a single field variable (as in Gross–Pitaevskii equation \cite{P-527}) to represent BEC in a trap, that shows decent qualitative matches with experimental data \cite{savage2003bose,albiez2005direct}. In an experimental work with double-well BEC in the presence of noise \cite{gati2006primary}, the theoretical modeling has been done using a two-mode BEC, analogous to what we are doing. 

In Sec. \ref{TSM}, we present a Bloch vector description of the system, and in Sec. \ref{ito}, the inclusion of Gaussian white noises into both the coupling strength and the detuning term is discussed. 

\subsection{Bloch vector description }\label{TSM}
We adopt a Bloch vector representation analogous to that used in spin systems. The relevant basis states are the fully molecular state and the fully atomic state, which are mapped onto the North and South poles of the Bloch sphere, respectively in Fig. (\ref{bloch sphere}). We define the components of the Bloch vector, also known as Schwinger pseudo-spin operators \cite{P70,P189}, as follows \cite{bloch4, commutator}:
$\hat{L}_x= \sqrt{2}(\hat{a}^\dagger \hat{a}^\dagger \hat{b} + \hat{b}^\dagger \hat{a} \hat{a})/N^{3/2}$, $\hat{L}_y=\sqrt{2}i(\hat{a}^\dagger \hat{a}^\dagger \hat{b} - \hat{b}^\dagger \hat{a} \hat{a})/N^{3/2}$, $\hat{L}_z=(2\hat{b}^\dagger \hat{b}-\hat{a}^\dagger \hat{a})/N$, and $N=2\hat{b}^\dagger \hat{b}+\hat{a}^\dagger \hat{a}$  

In this formulation, $\hat{L}_x$ and $\hat{L}_y$ correspond to the real and imaginary components of the coherence between the atomic and molecular states, respectively, while $\hat{L}_z$ quantifies the population imbalance or polarization between atoms and molecules \cite{cui2012atom}. Here $N$ is the total number of atoms in the system. Note that, the dynamics can be described using (i) $\hat{a},\hat{b}$ \cite{anglin2001dynamics}, (ii) Bloch components \cite{similar_hamiltonian3,similar_hamiltonian4}, or (iii) population imbalance and relative phase \cite{similar_hamiltonian3,BJJ20}. All are equivalent, but (ii) is most convenient, directly relating to observables. Consequently, MF and BBR analyses of two-mode systems are usually expressed via Bloch components. The relevant commutation relations among the Bloch vector components, which are essential for analyzing the dynamics of system, are listed in \cite{bloch4,liu2010shapiro}. It should be noted that for condensates in a double-well \cite{Linblad_master_equation}, the Bloch-vector components obey a closed $\mathrm{SU}(2)$ algebra \cite{stochastic_bosonic_josephson_junction} similar to conventional spin systems. In contrast, in the two-mode atom-dimer model, the Bloch-vector components do not satisfy the standard $\mathrm{SU}(2)$ algebra and obey an $\mathrm{SU}(1,1)$ algebra \cite{khripkov2011quantum} instead. This arises from the constraint in the large-$N$ limit, namely
$s_x^2 + s_y^2 = (1 + s_z)(1 - s_z)^2/2$
which indicates that the system's dynamics evolves on a generalized Bloch sphere \cite{bloch4}.
 The Bloch-vector description here is thus just a tool for analyzing the dynamics, and facilitating visualization.

\subsection{Inclusion of Noise}\label{ito}
We introduce stochastic noise terms that perturb the coupling strength $g$ and detuning $\epsilon_b$. These noise processes are assumed to have zero mean,  and are delta correlated \cite{noise_property}. Generally, noise in atomic–molecular BECs are attributed to collisions with thermal atoms 
\cite{bloch4,cui2012atom}. In this framework, condensed particles  constitute the system, while non condensed (thermal) particles act as the surrounding bath \cite{anglin1997cold,ruostekoski1998bose}, for which the Gaussian white-noise approximation is perfectly justified \cite{burt1997coherence}. The condensed mode  scatters off noncondensed thermal atoms, leading to a phase diffusion with energy $\gamma_x$ that depends on the thermal cloud temperature: acting as a noise on the Feshbach coupling. It has also been theoretically predicted \cite{savard1997laser} and experimentally demonstrated \cite{gardiner2000evaluation} that laser intensity fluctuations and
beam-pointing fluctuations can generate controlled noise, leading to heating and decoherence. The noise in detuning, on the other hand, stems from fluctuations in the applied magnetic field near the Feshbach resonance \cite{bloch_vector4} and from thermal fluctuations in the BEC \cite{saha2023phase}.
Defining $N{u_1}/V=U_1$, $N{u_2}/V=U_2$, $N{u_3}/V=U_3$ and $g\sqrt{N/V}=\tilde{g}$, Eq. (\ref{hamiltonian}) in the presence of this noise terms becomes in the large N limit:
 \begin{equation}
\label{hamiltonian_large_N}
\begin{split}
\hat{\mathcal{H}} = \frac{\hat{H}}{N} 
= & \frac{U_1}{8}(\hat{L}_z - 1)^2
+ \frac{U_2}{32}(\hat{L}_z + 1)^2 \\
& - \frac{U_3}{8}(\hat{L}_z^2 - 1)
+ \frac{\tilde{g} + n_x}{\sqrt{2}} \hat{L}_x
+ \frac{\epsilon_b + n_z}{4}(\hat{L}_z + 1)
\end{split}
\end{equation}
 Here $n_x$ and $n_z$ represent of noises in the Feshbach coupling and the
detuning respectively. The noises can be expressed as
$n_i(t)=\mathrm{d}w_i/\mathrm{d}t$, $w_i(t)$ ($i\in\{x,z\}$) being independent Wiener processes. The increments obey 
$\langle \mathrm{d}w_i\,\mathrm{d}w_j \rangle=\gamma_i\,\delta_{ij}\,\mathrm{d}t/2$ \cite{stochastic_bosonic_josephson_junction}.
 
 The large $N$ limit is completely justified because usual BEC experiments are performed with $10^5$-$10^6$ particles in magneto-optical traps \cite{strecker2003conversion,burt1997coherence}. 

Here, we consider only classical noise arising from external fluctuations, not quantum noise (the noise terms are not expressed as operators).  Consequently, our model includes fluctuations and decoherence, but excludes dissipation (particle loss). It is a closed system because the total number of particles ($\dot{N}=0$) is conserved, as shown in Appendix \ref{closed}. In Appendix \ref{dissipation}, we justify our choice of model in details, noting that two-, and three-body loss processes are not very significant for the present system.

In the next section (Sec. \ref{Relaxation dynamics thermodynamic quantum}), we explore how the Bloch vector components relax in both the thermodynamic (mean-field) and quantum (Bogoliubov backreaction) frameworks.

\section{Dynamical equations}\label {Relaxation dynamics thermodynamic quantum}
In Sec. \ref{MF_BBR}, we analyze the mean-field (MF) and beyond-mean-Field dynamics of the Bloch vector components. In Sec. \ref{clasiical analogy}, we establish an analogy between the MF and BBR regimes of the two-mode condensate,  with the weak and strong coupling limits of a classical spring-mass system. In Sec. \ref{mathematics}, we show that our system is analogous to a forced, damped harmonic oscillator. The frequency of oscillation ($\omega_i$), the intrinsically generated force ($F_i$), and the damping coefficient ($\lambda_i$) depend on the first (mean values) and second moments (variances and covariances) of the Bloch vector components.

\subsection{Mean field and Bogoliubov backreaction}\label{MF_BBR}
In this section, we focus on the relaxation dynamics of the Bloch vector components. We begin by analyzing the system using the MF approach, which involves only the first moments of the Bloch vector components. We then extend our analysis to include the effects of non-zero variance and covariance, capturing higher-order correlations through the BBR method.
Under the Gaussian (coherent-state) approximation, operator products are factored as
$\langle \hat{L}_i \hat{L}_j \rangle = \langle \hat{L}_i \rangle \langle \hat{L}_j \rangle$ \cite{Linblad_master_equation}.
Defining $s_i = \langle \hat{L}_i \rangle$ ($i = x, y, z$), we go beyond the MF limit using the Bogoliubov Born Green Kirkwood Yvon (BBGKY) hierarchy , which includes higher-order moments to capture correlations and quantum fluctuations. Specifically, a second-order truncation of the BBGKY method, termed Bogoliubov-back-reaction (BBR), is employed, where three operator expectations are approximated as:\
\begin{equation}
\begin{split}
   \label{higherarchy}
    \langle \hat{L}_i\hat{L}_j\hat{L}_k\rangle&\approx \langle \hat{L}_i\hat{L}_j\rangle \langle \hat{L}_k\rangle+\langle \hat{L}_i\rangle+\langle \hat{L}_j\hat{L}_k\rangle+\langle \hat{L}_i\hat{L}_k\rangle \langle \hat{L}_j\rangle\\&-2\langle \hat{L}_i\rangle \langle \hat{L}_j\rangle \langle \hat{L}_k\rangle 
\end{split}
\end{equation}

Moreover, we define the two-point correlation function $\Delta_{ij}$ as follows:\
\begin{equation}
    \label{normal correlation}
    \Delta_{ij}=\langle \hat{L}_i\hat{L}_j+\hat{L}_j\hat{L}_i\rangle-2\langle \hat{L}_i\rangle \langle \hat{L}_j\rangle
\end{equation}
Since Wick’s theorem \cite{chakrabarti1996quantum} allows higher-order moments to be factorized, we truncate the third-order moments in terms of the first- and second-order moments in Eq. (\ref{higherarchy}). For large $N$, with $\Delta_{ij} \sim N^{-1}$ and $r$-th cumulants $\sim N^{-(r-1)}$, higher-order correlations become negligible. Moreover, $\Delta_{ij}(t)$ relaxes over time. Hence, the BBGKY hierarchy effectively truncates to the BBR level \cite{split_correlation,anglin2001dynamics}, as detailed in Appendix \ref{molecule}. 

When $\Delta_{ij} \neq 0$, the system lies in the beyond-MF (BBR) regime, reflecting the presence of nontrivial correlations. In contrast, $\Delta_{ij} = 0$ corresponds to the MF regime, where the system evolves within the eigenstates of the relevant observable \cite{P33}, and no correlations are present. Equivalently, vanishing variance or covariance indicates that the observable coincides with its mean value, signifying that it is an eigenstate of the system. By incorporating both first- and second-order moments in the large-$N$ limit, the system dynamics can be constructed as \cite{mukherjee2025relaxation}:
\begin{subequations}
\label{covariance}
    \begin{equation}
    \label{covariance1}
    \begin{split}
    \dot{s_x}=c_1(\Delta_{yz}+2s_ys_z)+c_2s_y-\frac{\gamma_z s_x}{2}
         \end{split}
    \end{equation}
    \begin{equation}
    \label{covariance2}
\begin{split}
    \dot{s_y}=&-c_1(\Delta_{zx}+2s_zs_x)-c_2s_x-\frac{\tilde{g}}{\sqrt{2}}(1+2s_z-\frac{3}{2}\Delta_{zz}-3s^2_z)\\&+\gamma_x(-2s_y+3\Delta_{yz}+6s_ys_z)-\frac{\gamma_z s_y}{2}
        \end{split}
    \end{equation}
    \begin{equation}
    \label{covariance3}
    \begin{split}
    \dot{s_z}=2\sqrt{2}\tilde{g} s_y-\gamma_x(1+2s_z-\frac{3}{2}\Delta_{zz}-3s^2_z)
       \end{split}
    \end{equation}
    \begin{equation}
\begin{split}
    \dot{\Delta}_{xx}=4c_1s_y\Delta_{zx}+2(2c_1s_z+c_2)\Delta_{xy}-\gamma_z\Delta_{xx}
        \end{split}
        \end{equation}
        \begin{equation}
\begin{split}
    \dot{\Delta}_{yy}=&\bigg(-4c_1s_x-2\sqrt{2}\tilde{g} (1-3s_z)+12\gamma_xs_y\bigg)\Delta_{yz}\\&-2(2c_1s_z+c_2)\Delta_{xy}-\bigg(\gamma_z+4\gamma_x(1-3s_z)\bigg)\Delta_{yy}
        \end{split}
        \end{equation}
        \begin{equation}
\begin{split}
    \dot{\Delta}_{zz}=4\sqrt{2}\tilde{g}\Delta_{yz}-4\gamma_x (1-3s_z)\Delta_{zz}
        \end{split}
        \end{equation}
        \begin{equation}
\begin{split}
    \dot{\Delta}_{xy}=&-(2c_1s_z+c_2)\Delta_{xx}+(2c_1s_z+c_2)\Delta_{yy}+2c_1s_y\Delta_{yz}\\&+\bigg(6\gamma_xs_y-2c_1s_x-\sqrt{2}\tilde{g}(1-3s_z)\bigg)\Delta_{zx}\\&-\bigg(\gamma_z+2\gamma_x(1-3s_z)\bigg)\Delta_{xy}
        \end{split}
        \end{equation}
        \begin{equation}
\begin{split}
\dot{\Delta}_{yz}=&\bigg(-2c_1s_x-\sqrt{2}\tilde{g}(1-3s_z)+6\gamma_x s_y\bigg)\Delta_{zz}-(2c_1s_z+c_2)\Delta_{zx}\\&+2\sqrt{2}\tilde{g}\Delta_{yy}-\bigg(\frac{\gamma_z}{2}+4\gamma_x(1-3s_z)\bigg)\Delta_{yz}    
\end{split}
        \end{equation}
  \begin{equation}
\begin{split}
    \dot{\Delta}_{zx}=&2c_1s_y\Delta_{zz}+(2c_1s_z+c_2)\Delta_{yz}+2\sqrt{2}\tilde{g}\Delta_{xy}\\&-\bigg(2\gamma_x(1-3s_z)+\frac{\gamma_z}{2}\bigg)\Delta_{zx}
        \end{split}
        \end{equation}
        \end{subequations}
 Here, $c_1 =
U_3/2\hbar-U_1/2\hbar-U_2/8\hbar$, and $c_2=U_1/\hbar-U_2/4\hbar-\epsilon_b/\hbar$. Both $c_1$ and $c_2$ are functions of the fundamental experimental parameters, namely $u_1$, $u_2$, $u_3$, and $\epsilon_b$. In the BBR case, the relaxation times may increase or decrease relative to the relaxation times found from MF, depending on the parameters $|s_z^0|$ and $\epsilon_b$.

The first and second moments of all three components of the Bloch vector are presented in terms of molecular fractions in Appendix. \ref{molecule}, assuming the system is in a Fock state. It is to be noted that the Eq. (\ref{covariance}) would have been structurally similar to the ones obtained in \cite{bloch4} employing Lindblad Formalism to study dephasing, if we considered noise in the coupling only (and not in the Feshbach detuning) . 
\subsection{Classical analogy}{\label{clasiical analogy}}
 Consider a classical system consisting of three springs and two bobs, where each bob is connected to its respective stand via a spring. In addition, the two bobs are interconnected by an intermediate spring. If the spring constant of the intermediate spring is equal to that of the two boundary springs, the system is in the strong coupling limit, where energy is rapidly transferred between the bobs. In contrast, if the intermediate spring has a much lower spring constant than the boundary springs, energy transfer between the bobs occurs more slowly, known as the weak coupling limit.
 
 For a resonantly-coupled system of atomic-molecular BECs, these strong and weak coupling limits correspond to the BBR, and MF conditions, respectively. In weak coupling regime, the modes are nearly independent, structurally resembling the mean-field theory scenario. A stronger coupling enhances mode mixing and promotes the buildup of correlations, more suited for beyond-mean-field descriptions such as BBR.

 If the system is taken out-of-equilibrium, a weakly-coupled spring-bob system relaxes slower due to the slow energy-transfer processes. For a strong-coupled system, there is efficient energy transfer among different modes, and the system relaxes faster. We shall see in  Sec \ref{long-trans} that this indeed is the case with MF and BBR : inclusion of correlations leading to a shorter relaxation time in the later. 
 
 However, the picture changes near the resonance. Now the two modes have nearly matched energies, and a stronger coupling actually leads to a more persistent oscillation: and longer relaxation time. The weakly coupled system, on the other hand, relaxes relatively faster near resonance, when compared to the strongly coupled one. Again, this trait is also captured in the MF vs. BBR comparison in Sec \ref{long-trans}: where the relaxation time corresponding to BBR shoots up near the zero-detuning value. 
 
 \begin{figure}[h]
    \centering
    \includegraphics[width=0.6\linewidth]{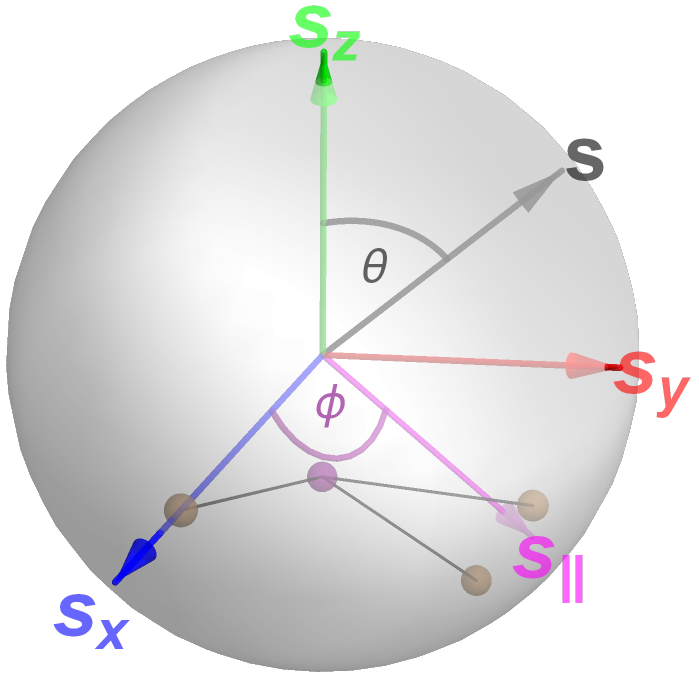}
    \caption{A few initial states (brown) and the final equilibrium point (purple, south pole oriented state) are indicated on the Bloch sphere, where $\theta$ and $\phi$ denote the polar and azimuthal angles, respectively.}
    \label{bloch sphere}
\end{figure}
\begin{figure}[h]
\centering
    \includegraphics[width=0.8\linewidth]{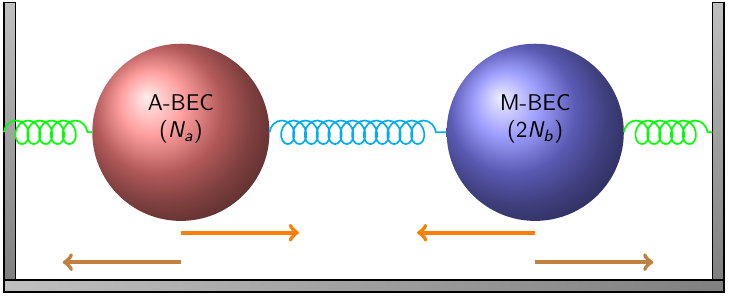}
\caption[short description]{Two types of motions for A-BEC ($N_a$) and M-BEC ($2N_b$).}
\label{out of phase}
\end{figure}
\subsection{Nature of fluctuations}{\label{mathematics}}
By taking higher-order time derivatives of Eqs. (\ref{covariance1}), (\ref{covariance2}), and (\ref{covariance3}), we obtain a set of coupled, forced-damped oscillatory equations \cite{P14, P51} governing the dynamics of all Bloch vector components $s_i$.

 \begin{equation}
\label{EOM}
\ddot{s}_i+\lambda_i\dot{s}_i+\omega^2_i s_i = F_i 
\quad \text{while} \quad i \in \{x,y,z\}
\end{equation}

\noindent
The detailed expressions of  the damping rates ($\lambda_i$), frequencies ($\omega_i$), and forces ($F_i$) are placed in  Appendix \ref{appendixA}. We observe that $F_i$ are intrinsically generated forces, not external drives, since they depend only on the first and second moments of the Bloch-vector components.

The effective frequency ($\omega^{\text{eff}}_i$) of under-damped ($\lambda_i/\omega_i<1$\cite{P51}) force oscillation can be formed from Eq. (\ref{EOM}), 
\begin{equation}
\label{effective frequency}
  \omega^{\text{eff}}_i=\sqrt{\omega^2_i-\frac{\lambda^2_i}{4}}  
\end{equation}
 
\begin{figure}
\centering
\begin{subfigure}{0.8\linewidth}
    \centering
    \includegraphics[width=\linewidth]{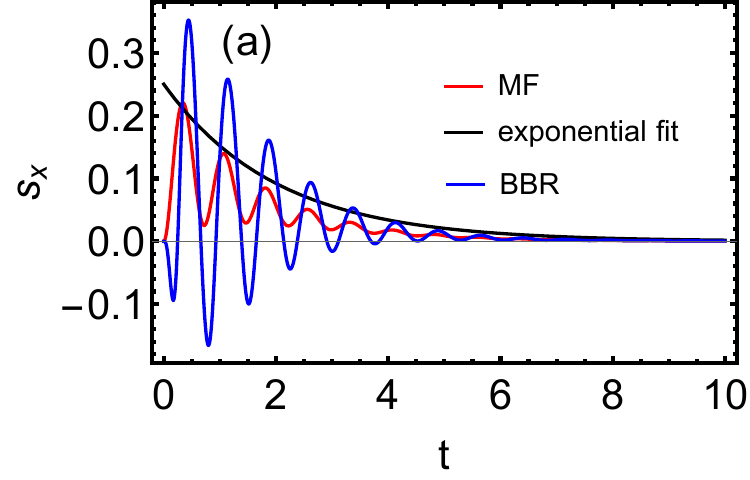}
    \phantomcaption 
    \label{relaxation s_x}
\end{subfigure}
\begin{subfigure}{0.8\linewidth}
    \centering
    \includegraphics[width=\linewidth]{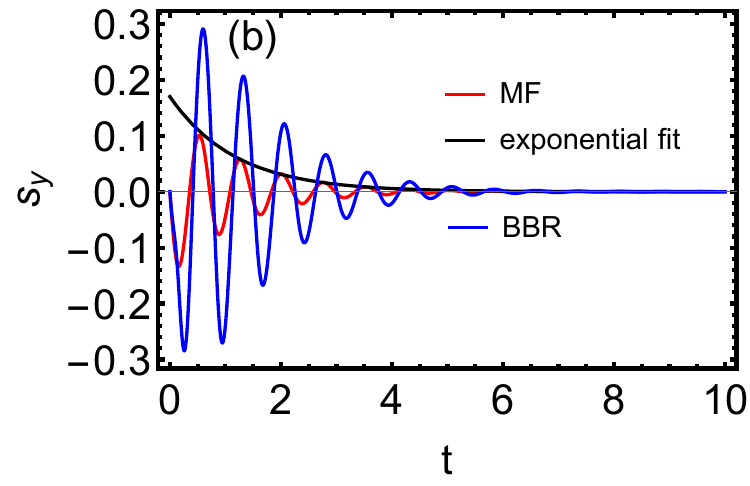}
    \phantomcaption 
    \label{relaxation s_y}
\end{subfigure}
\begin{subfigure}{0.8\linewidth}
    \centering
    \includegraphics[width=\linewidth]{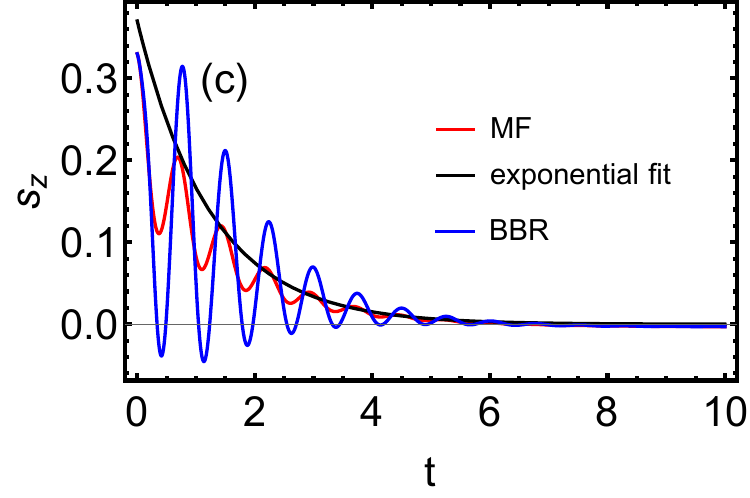}
    \phantomcaption 
    \label{relaxation s_z}
\end{subfigure}
\caption[short description]{Relaxation dynamics of (a) $s_x$, (b) $s_y$, and (c) $s_z$, where the mean-field (MF) case ($\Delta^0_{ij}=0$), the BBR case ($\Delta^0_{ij}\neq 0$), and the exponential fit are distinguished by color, with the Feshbach detuning set to $\epsilon_b = 9$.}
\label{relaxation s}
\end{figure}
\begin{figure}[h]
\centering
    \includegraphics[width=0.8\linewidth]{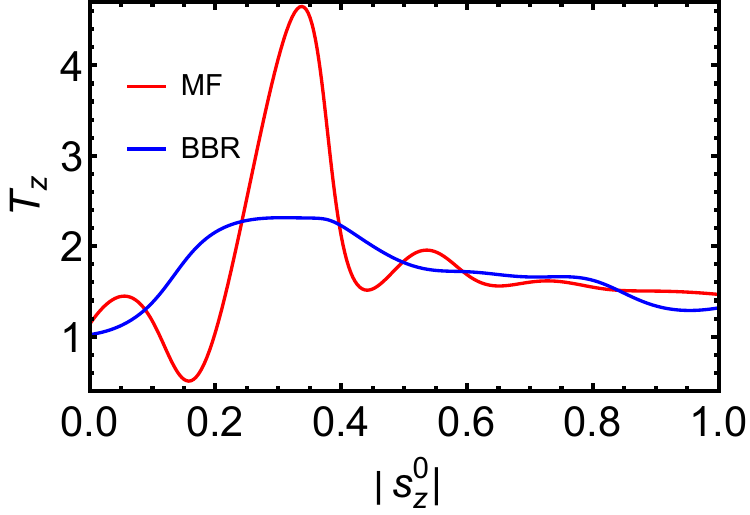}
    \caption[short description]{
Effect of absolute initial polarization ($\lvert s^0_z\rvert$) variation with longitudinal relaxation time ($T_{z}$) where MF (\red), and BBR (\blue) are denoted by their respective colors with the Feshbach detuning set to $\epsilon_b = 9$.
}
\label{longitudinal}
\end{figure}

In Sec. \ref{long-trans}, we show that the oscillation profiles of the Bloch vectors indeed relax resembling a forced-damped oscillator. Before that, in  Sec. \ref{energy or particle transfer}, we specify the initial conditions of the system and the experimental parameters used to study its dynamics.
\section{Choice of Parameters and Initial Conditions}{\label{energy or particle transfer}}
 
        In Fig. (\ref{bloch sphere}), the initial state is indicated by a brown ball, whereas the final equilibrium state, depicted by a purple atom-heavy ball. The quantities $s$ and $s_\parallel$
	denote the Bloch vector and its projection onto the equatorial plane, respectively. 
    
   For numerical solutions, we consider the physical parameters corresponding to ${}^{87}\mathrm{Rb}$ condensates. 
This is because experiments on atomic--molecular Bose condensates have been performed with ${}^{87}\mathrm{Rb}$ in the past 
\cite{papp2006observation,zhang2021transition}, and in a recent experiment with ${}^{87}\mathrm{Rb}$ condensate it was 
observed that the relative phase and atom-number imbalance relax to a phase-locked steady state \cite{BJJ1}. 
The system parameters are thus assigned as follows: 
$U_1:U_2:U_3:\tilde{g}:\epsilon_b = 1:4.3:-7:2:9.1$. 
In our case, these translate to $c_1 = -4.5$ and $c_2 = -9$.
 
 In our model, the signal-to-noise ratio (SNR) for both the coupling noise ($\gamma_x$) and the detuning noise ($\gamma_z$) are taken to be $\approx 10$. We demonstrate in Appendix \ref{initial state} that such a choice falls well within experimentally viable range. The choice of this particular SNR is motivated by studies on stochastic resonance in ultracold atomic systems \cite{witthaut2008dissipation, wellens2003stochastic} where a noise of $10\%$ on top of the original amplitude was found to impact the relaxation process substantially and yet not destroy the  core system dynamics. If the noise strength is weak enough, the system is expected to remain in purely MF-domain, and a $10\%$ noise (as in the stochastic resonance studies) is actually a good platform to study beyond-mean-field effects. 
 
The notion of noise with comparable relative strength is, in fact, relevant across a variety of ultracold-atom experiments operating near resonance. For instance, in the kicked-rotor experiment of \cite{sadgrove2005effect}, quantum resonance peaks were found to remain robust even in the presence of strong amplitude noise in the kicking strength, with fluctuations ranging from zero up to twice the mean value. In addition, in a two-mode Bose–Einstein condensate experiment on noise interferometry \cite{gati2006noise}, the relative phase fluctuations were on the order of $\sim 0.3$–$1$ rad, corresponding to an effective fluctuation level of roughly $5-15\%$ when normalized by $2\pi$. These examples indicate that the noise strengths considered in our work lie well within experimentally relevant regimes. 

 The choice of initial conditions, evaluation of experimental parameters, and adopted noise terms are presented in Appendix \ref{initial state}.

        In Sec \ref{long-trans}, we demonstrate the behaviour of both longitudinal and transverse relaxation time as a function of initial polarization ($\lvert s^0_z\rvert$) and Feshbach detuning ($\epsilon_b$).
        
\begin{figure}[h]
\includegraphics[width=0.8\linewidth]{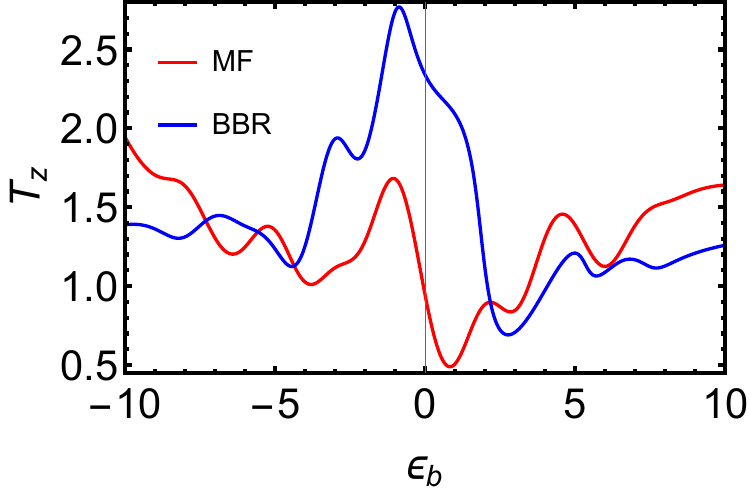}
\caption[short description]{Longitudinal relaxation time ($T_z$) as a function of the detuning ($\epsilon_b$) where, MF (\red) and BBR (\blue) denote the corresponding results, with the initial polarization set to $s^{0}_z=0$.}
\label{longitudinal_relax}
\end{figure}
\section{Relaxation time}\label{long-trans}
 In Sec. \ref{stability analysis}, we identify the stable equilibrium point of the dynamics through the Jacobian matrix formalism. In Sec. \ref{just definition}, we evaluate both $T_z$ and $T_{x\text{-}y}$, where $\lvert s^0_z\rvert$, and $\epsilon_b$ are treated as variables.
\subsection{Stable Equilibrium}\label{stability analysis}
In the absence of noise ($\gamma_x=\gamma_z=0$), if $\Delta_{ij}=0$, then from Eq. (\ref{covariance3}) we obtain $s_y=0$ when $\dot{s}_z=0$. Incidentally, $s_y=0$ also makes $\dot{s}_x=0$ in Eq. (\ref{covariance1}). Furthermore, substituting $s_y = 0$ and $\dot{s}_y = 0$ into Eq. (\ref{covariance2}) yields the condition $(2c_1 s_z + c_2) \neq 0$, which leads to the solutions $s_z = 1$ or $s_z = -1/3$. For these values of $s_z$, we obtain $s_x=-\tilde{g}(s_z-1)(s_z+1/3)/\sqrt{2}(2c_1s_z+c_2)$\\
Since, the Jacobian can be defined as \cite{greiner2003classical},

\begin{equation}
\label{matrix}
  J_{ij}=\frac{\partial s_i}{\partial s_j}  
\end{equation}
So, numerical solutions of Eq. (\ref{matrix}) leads to an unstable equilibrium point, $(0,0,1)$ with
 \begin{subequations}
  \begin{equation}
  \label{unstable eigen}
  \begin{split}
&\lambda_1=-0.22-10.56i, \quad   \lambda_2=-0.22+10.56i \\&\text{and},\quad \lambda_3=1.14,      
  \end{split}  
  \end{equation}
  \text{and a stable equilibrium point $(0,0,-1/3)$ with}
  \begin{equation}
  \label{stable eigen}
  \begin{split}
 &\lambda_1=-0.882-13.26i, \quad   \lambda_2=-0.882+13.26i\\&\text{and},\quad \lambda_3=-0.737     
  \end{split}
  \end{equation}
 \end{subequations}
 The detailed derivations are presented in Appendix \ref{appendix B}.\\
 Now, the real parts of the complex-conjugate eigenvalues are negative in both Eqs. (\ref{unstable eigen}) and (\ref{stable eigen}); therefore, both cases exhibit an inward spiral \cite{strogatz1994nonlinear}. However, Eq. (\ref{unstable eigen}) has one positive real eigenvalue, $\lambda_3$, which causes the system to diverge along that direction. In contrast, $\lambda_3$ in Eq. (\ref{stable eigen}) is negative, so the system converges in all directions and yields a stable solution. Furthermore, the equilibrium point $\left(0,0,-1/3\right)$ is found to be stable (an attractor), indicating that the atom-heavy state is dynamically preferred, which is denoted by $s^{\text{eq}}_z$. The condition $s_z = -1/3$ corresponds to equal populations of A-BEC and M-BEC. So it is this configuration that the system eventually arrives at. 
 
\subsection{Longitudinal and transverse relaxation time}{\label{just definition}}
Fig. (\ref{relaxation s}) shows the relaxation dynamics of the oscillation envelope as the system starts from $s_i=0$ ($i\in\{x,y,z\}$). The relaxation process is studied using both MF and BBR methods \cite{marino1999bose, saha2023phase}. 
We assume an exponential decay governed by
\begin{equation}
\lvert s_i(t)\rvert = \lvert s^0_i\rvert e^{-t/\tau},
\end{equation}
where $\tau$ is the decay time constant and 
$s^0_i$ is the initial amplitude. Such an exponential decay typically arises when the system time scale is comparable with the noise time scale.  The time interval between two measurements is
$\Delta t = t_{\text{f}} - t_{\text{in}}$, where $t_{\text{f}}$ and $t_{\text{in}}$ are the final and initial times, respectively. The exponential relaxation time is then approximated as
\begin{equation}
\label{exponential fall}
\tau_{\text{relax}}
= \frac{\Delta t}{\bigg\lvert\ln\left( \dfrac{ s_i^0}{ s_i} \right)\bigg\rvert} .
\end{equation}

The system relaxes to the equilibrium configuration $\left(0,0,-1/3\right)$ as obtained from the MF equations in Sec. \ref{stability analysis}, for which we shift both $s_z^0$ and $s_z$ by $+1/3$. Fig. \ref{relaxation s} shows that, even with correlations included through the BBR equations, the system relaxes to the same equilibrium. The blue curves correspond to exponential fits, with damping rates evaluated from Eq.(\ref{decay rate}) at $s_z = 0$ (zero imbalance). We approximate $s_x \approx A_x e^{-t/T_x}$, $s_y \approx A_y e^{-t/T_y}$, and $s_z \approx A_z e^{-t/T_z}$, where $A_x$, $A_y$, and $A_z$ are the amplitudes. For these plots, $A_x=0.25$, $A_y=0.17$, and $A_z=0.37$, reflecting particle leakage from the M-BEC to the A-BEC state.

   We also compute the relaxation times $T_z$ (relaxation time for population-imbalance)  and $T_{x\text{-}y}$ (relaxation time for coherence) for varying initial polarization ($\lvert s^0_z\rvert$) and detuning ($\epsilon_b$). The effective decay rate in the anisotropic transverse ($x\text{-}y$) plane is best characterized by the harmonic mean \cite{P61, P211}:
\begin{equation} \label{harmonic mean}
T_{x\text{-}y} = \frac{2T_x T_y}{T_x + T_y}
\end{equation}.
  
Fig. (\ref{longitudinal}) and Fig. (\ref{longitudinal_relax}) are about the relaxation of the population imbalance, as the initial imbalance and detuning are varied respectively. In Fig. (\ref{longitudinal}), it is shown that as the imbalance approaches $s^{\text{eq}}_z$, which is the stable point of the dynamics, the relaxation rate reaches a minimum and $T_z$ attains its maximum. Here, the MF supports small persistent oscillations near $s^{\text{eq}}_z$, and the damping is weak. The BBR, however, kills the oscillations faster by introducing damping via fluctuations. As a result, the peak height of $T_z$ is higher for MF, compared to BBR. Away from the equilibrium point, both MF and BBR follow similar decay channels. 

The transition probability between the atomic and molecular states is maximal, leading to a maximum relaxation rate and, consequently, a minimum $T_z$ near the resonance point, as shown in Fig. (\ref{longitudinal_relax}). It is to be noted that the resonance point actually shifts by a small amount along the positive detuning side: as explained in Sec. \ref{physical insight}. In this regime, interactions become stronger and 
fluctuations grow : something that is not captured well in MF, and hence it predicts a simple and almost instantaneous relaxation. The BBR, however, accommodates longer-lived correlations, leading to a higher (and more physical) value of $T_z$.
 
 \begin{figure}[H]
\centering
    \includegraphics[width=0.8\linewidth]{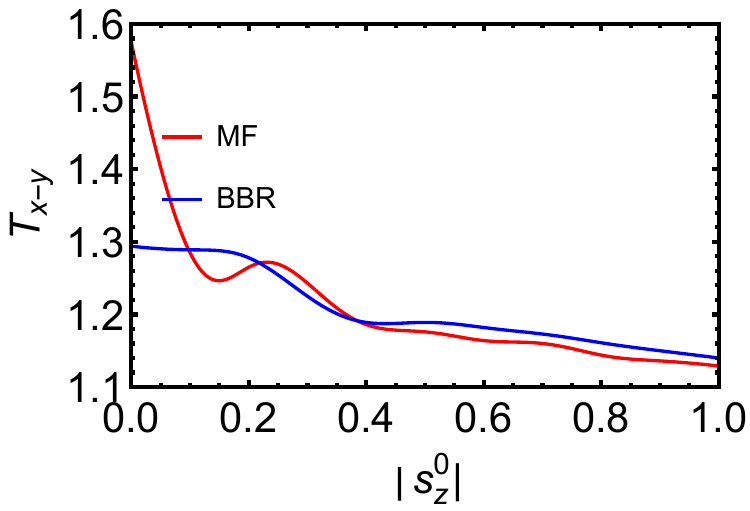}
    \caption[short description]{
Effect of absolute initial polarization ($\lvert s^0_z\rvert$) variation with transverse relaxation time ($T_{x\text{-}y}$) where MF (\red), and BBR (\blue) are denoted by their respective colors,  with the Feshbach detuning set to $\epsilon_b = 9$.
}
\label{transverse}
\end{figure}
\begin{figure}[h]
\includegraphics[width=0.8\linewidth]{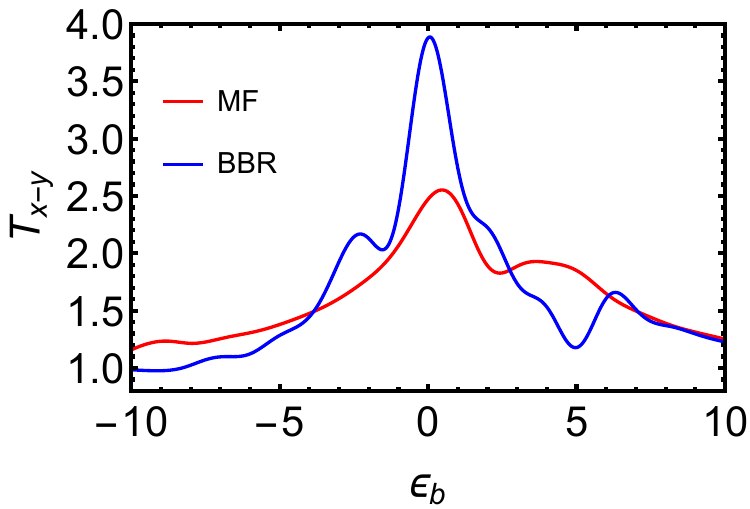}
\caption[short description]{Effect of variation of detuning ($\epsilon_b$) on transverse relaxation time ($T_{x\text{-}y}$) where MF (\red), and BBR (\blue) are denoted by their respective colors, with the initial polarization set to $s^0_z = 0$.
}
\label{coherence_relaxation_time}
\end{figure}
Fig. (\ref{transverse}) and Fig. (\ref{coherence_relaxation_time}) depict the relaxation of coherence with varying initial imbalance and detuning. In Fig. (\ref{transverse}), an enhanced $\lvert s^0_z\rvert$ takes the system away from the equatorial plane of the Bloch sphere, leading to a reduced value of coherence. As a result, the coherence itself decays faster, allowing $T_{x\text{-}y}$ to decrease. As expected, BBR deviates strongly from MF near the equator $\lvert s^0_z\rvert=0$, and the difference gradually diminishes as system moves towards higher $\lvert s^0_z\rvert$ and lower-coherence regimes.

In Fig. (\ref{coherence_relaxation_time}), it is evident that $T_{x\text{-}y}$ shows a maximum near resonance, both for MF and BBR. This is because the fluctuations are strongly amplified near the resonance, and the coherence relaxation is delayed (more so, when non-zero variances and covariances are incorporated via BBR). Away from the resonance point, both MF and BBR relaxes in the nearly similar fashion. 

\begin{figure}[h]
\centering
    \includegraphics[width=0.8\linewidth]{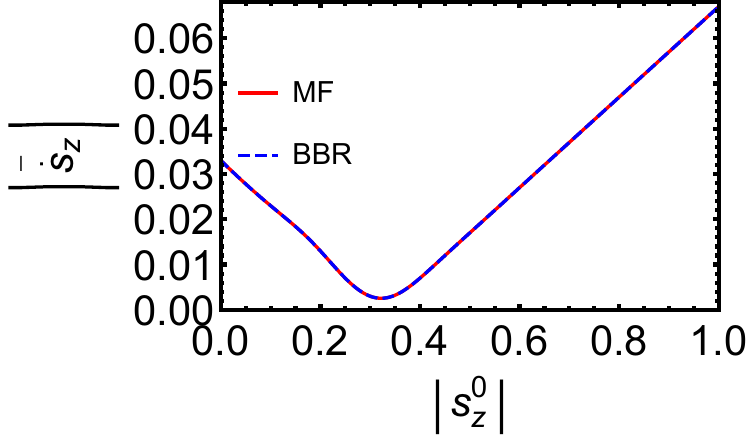}
\caption[short description]{Time-averaged drift speed ($\lvert\bar{\dot{s}}_z\rvert$) influenced by the absolute initial imbalance ($\lvert s^0_z\rvert$) where, 
MF (\red), and BBR (\bluedashed) denoted by their respective colors.}
\label{drift}
\end{figure}
 
In Sec. \ref{intution}, we employ several markers as supporting circumstantial evidences to study the behavior of $T_z$ and $T_{x\text{-}y}$.

\section{ Relaxation Times: as Captured in other physical quantities}\label{intution}
The time-averaged expectation of any operator $\hat{A}$ is defined as \cite{mukherjee2025relaxation}
\begin{equation}
\langle \hat{A} \rangle_t = \bar{A} =
\frac{\int_{t_i}^{t_f} A(t)dt}{\int_{t_i}^{t_f} dt}
\end{equation}
We compute several such quantities in the range $t_i = 0$ to $t_f = 10$, the interval capturing the relaxation dynamics. Here, $t_i=0$ corresponds to the initial condition with equal population and zero coherence, and beyond $t_f=10$, fluctuations decay as the system approaches a steady atom-favored state (Fig. \ref{relaxation s}) : so any expectation value calculated in this range reflects the gross quantitative behavior along the relaxation path. 

It is to be noted that the system is not in equilibrium or steady state in this time-window, and technically, not in the ergodic domain. Still we use the time-averaging as a crude marker because it effectively characterizes the dominant system traits.

 We investigate the effects of $\lvert s^0_z\rvert$, and $\epsilon_b$ on a few relevant physical quantities and attempt to relate these findings to the trends observed in the relaxation process. Specifically, we focus on four key quantities: (i) the time-averaged drift speed ($\lvert \bar{\dot{s}}_z \rvert$) and (ii) the time-averaged conversion efficiency ($\bar{\Gamma}$), which characterize the behavior of $T_z$ as functions of the initial imbalance $\lvert s_z^0\rvert$ and the detuning $\epsilon_b$ (Sec. \ref{longitudinal marker}); and (iii) the time-averaged von Neumann entropy ($\bar{S}_{\mathrm{V}}$) and (iv) the time-averaged condensate fraction ($\bar{C}_+$), which probe the transverse dynamics $T_{x\text{--}y}$ under variations of $\lvert s_z^0\rvert$ and $\epsilon_b$ (Sec. \ref{transverse marker}).
 Subsequently, Sec. \ref{physical insight} provides the physical insight underlying the shift of the resonance position toward positive detuning (i.e., to the right of the resonance point).
\begin{figure}[h]
    \centering
    \includegraphics[width=0.8\linewidth]{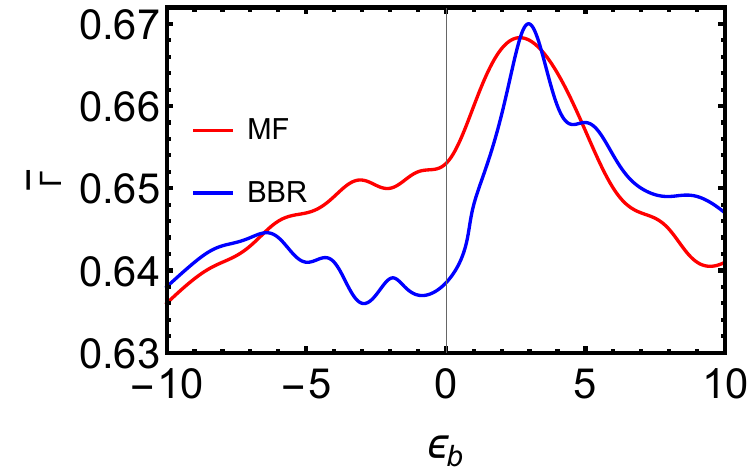}
\caption[short description]{The time-averaged conversion-efficiency ($\bar{\Gamma}$) is motivated by detuning ($\epsilon_b$), where the following methods are represented by their respective colors: MF (\red), and BBR (\blue).
}
\label{conversion efficiency}
\end{figure}

\subsection{Longitudinal relaxation time}{\label{longitudinal marker}}
\subsubsection{Drift speed: A Quantitative Marker of Longitudinal Relaxation Time under Variable Initial Imbalance}
An increase in $\lvert s^0_z\rvert$ toward the atomic state drives the system from an oscillatory regime into a gradually frozen domain. In this process, the magnitude of the drift speed $\lvert\bar{\dot{s}}_z\rvert$ initially decreases before increasing again, as illustrated in Fig. (\ref{drift}). This can be connected to the corresponding relaxation time $T_z$ attaining its maximum near the equilibrium point $(0,0,-1/3)$ (as in Fig. (\ref{longitudinal})), while decreasing on either side of this equilibrium.
This is because the lower the drift speed, the slower would be the relaxation mechanism.

\subsubsection{Probing Longitudinal Relaxation Time via Conversion Efficiency as a function of Feshbach detuning}

The Conversion efficiency is defined as\begin{equation}
\label{formula conversion}
    \Gamma=\frac{2N_b}{N}=\frac{1+s_z}{2}
\end{equation}
In Fig. (\ref{conversion efficiency}), $\bar{\Gamma}$ reaches its maxima near resonance, and as a result of this, $T_z$ reaches its minima. This is because the transition from one state to other takes a shorter time at this point, and the system relaxes fast. The peak of $\bar{\Gamma}$ appears at the same $\epsilon_b$ value for MF, and BBR.  
\begin{figure}[h]
    \centering
    \includegraphics[width=0.8\linewidth]{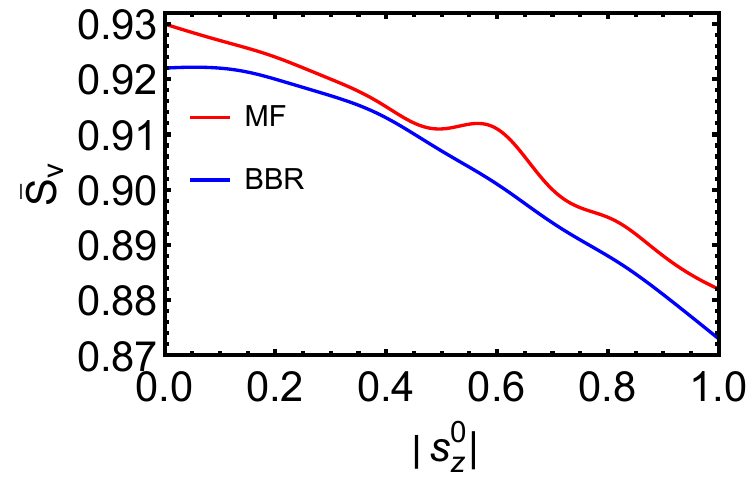}
\caption[short description]{The time-averaged von-neumann entropy ($\bar{S}_\text{V}$) as a function of absolute initial imbalances ($\lvert s^0_z\rvert$) where, MF (\red), and BBR (\blue)}
\label{entropy}
\end{figure}
\subsection{Transverse relaxation time}{\label{transverse marker}}
\subsubsection{Linking Von-neumann entropy to Transverse Relaxation under Variable Initial Imbalance}

The density matrix can be represented as \cite{wiener,P510}
\begin{equation}
\hat{\rho} = \frac{I + \sigma_j s_j}{2}, \quad \text{where} \quad j \in \{x, y, z\},
\label{density matrix}
\end{equation}

Here, $\sigma_j$ is ($2\times2$) pauli matrices. In matrix form, Eq. (\ref{density matrix}) takes the form:
\begin{equation}
\hat{\rho} = \frac{1}{2} \begin{pmatrix}
1 + s_z & s_x - i s_y \\
s_x + i s_y & 1 - s_z
\end{pmatrix}.
\label{eigen value}
\end{equation}\\
Now, purity is defined as \cite{P225}, 
\begin{equation} P=\text{Tr}{{\hat\rho}^2}=\frac{1+\sum_j s_j^2}{2}
\label{purity}
\end{equation} which is reduced through the decoherence process. Here, $P = 1$ corresponds to a pure state, while $P = 1/2$ corresponds to a maximally mixed state. Here, the diagonal elements represent atomic and molecular populations, while the off diagonal elements represent coherences.

The von Neumann entropy is given by \cite{P510}:
\begin{equation}
\begin{split} 
S_{\text{V}}(P)=&-\frac{1+\sqrt{2P-1}}{2}\log_2\left(\frac{1+\sqrt{2P-1}}{2}\right)\ \\&-\frac{1-\sqrt{2P-1}}{2}\log_2\left(\frac{1-\sqrt{2P-1}}{2}\right) \end{split} 
\label{entropy log} 
\end{equation} where $S_{\text{V}}(P) = 0$ and $S_{\text{V}}(P) = 1$ correspond to pure and maximally mixed states, respectively.\\
From Fig. (\ref{entropy}), we observe that $\bar{S}_{\text{V}}$ decreases as $\lvert s^0_z\rvert$ increases. The tendency of $\lvert s^0_z\rvert$ approaching a pure state enhances the loss of coherence, leading to a decrease in $T_{x\text{-}y}$ as in Fig. (\ref{transverse}). The evolution of the system toward a pure state (i.e., a fully atomic state) is also indicated by this reduction in $\bar{S}_V$. The overall trend is the same for MF as well as BBR. However, finer details like the exact nature of oscillations vary, as the governing equation (Eq. \ref{covariance}) are all nonlinear and dynamics extensively depend on initial conditions.

\subsubsection{Condensate fraction as an Indicator of Transverse Relaxation Dynamics as a function of Feshbach detuning}
\begin{figure}[h]
    \centering
    \includegraphics[width=0.8\linewidth]{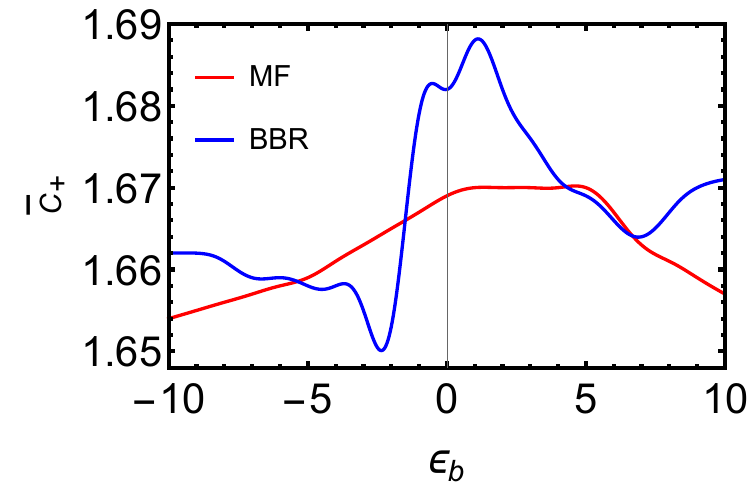}
\caption[short description]{
Time-averaged condensate part ($\bar{C}_+$) is controlled by detuning ($\epsilon_b$), where the following methods are represented by their respective colors:
 MF (\red), and BBR (\blue).
}
\label{condensate}
\end{figure}

From, Eq. (\ref{eigen value}), we obtains these eigen values
\begin{equation}
    C_\pm=\frac{1\pm\sqrt{s^2_z+4\{s^2_x+s^2_y+(1-s_z)(1+3s_z)\}}}{2}
\end{equation}  
Here, $C_\pm$ \cite{pudlik2013dynamics} denote the condensed and non-condensed components, respectively; since they are obtained from the diagonalization of the density matrix, they also contain information about coherence.  Near the resonance position the condensate fraction  $\bar{C}_+$ reaches its maximum for all cases, as shown in Fig. (\ref{condensate}).
Thus, near resonance, the coherence is highest as well, indicative of a longer relaxation time as shown in Fig. (\ref{coherence_relaxation_time}).

\subsection{Why do the Resonance points shifts to the Positive Detuning Side?}\label{physical insight}

\subsubsection{shifts due to magnetic field fluctuation}
 New resonance point ($B_{\text{res}}$) due to magnetic field fluctuation ($\delta B$) in $\epsilon_b$ is \cite{chin2010feshbach},
 \begin{subequations}
   \begin{equation}
B_0=B_{\text{res}}+\delta B     
 \end{equation}
 \text{From definition of $\epsilon_b$, one obtains,} 
 \begin{equation}
 \label{modified}
  \epsilon_b= \epsilon^0_b-\mu_{\text{co}}\delta B,\quad\text{where}\quad  \delta B=-\frac{\delta \Delta B}{\delta\mu_{\text{co}}}  
 \end{equation} Here, $\epsilon^0_b=\mu_{\text{co}} (B-B_\text{res})$ where $B$ and $\mu_{\text{co}}$ represent the applied magnetic field and the difference in magnetic moments between the closed and open channels, respectively.
 \text{So, From Eq. (\ref{modified}), we obtain \cite{P-527}}   
 \end{subequations}
\begin{subequations}
 \begin{equation}
\label{alpha}
  a(B)=\alpha g^2 \quad\text{where}\quad   a_{\text{eff}}(B)=a_{\text{bg}} (1-\frac{\Delta B}{B-B_0})
\end{equation}
\text{where $\alpha = m_{\text{a}}/4\pi \hbar^2 \Delta B \mu_{\text{co}}$; $m_a$, $a_{\text{bg}}$, and $\Delta B$ denote the}\\
{atomic mass, background scattering length, and resonance width respectively.}   
From, Eq. (\ref{alpha}) we obtain\\
 \begin{equation}
\label{shift}
    B-B_0=\Delta B (1-\alpha \tilde{g}^2)^{-1}
\end{equation}
\end{subequations}

Since, $\alpha \tilde{g}^2=\bigg(1-\Delta B/(B-B_0)\bigg)^2$, and for positive scattering \\length in BEC make, $\alpha \tilde{g}^2<1$. Now, Eq. (\ref{shift}), produces,

\begin{subequations}
\begin{equation}
 B-B_0=\Delta B(1+\alpha \tilde{g}^2)   
\end{equation}   
Now, Eq. (\ref{modified}), generates
\begin{equation}
    \delta B=-\Delta B(1+\alpha \tilde{g}^2), \quad \text{and}\quad
 B_{\text{res}}-B_0=\Delta B(1+\alpha \tilde{g}^2)=\text{+ve}   
\end{equation}    
\end{subequations}
So, the resonance point shifts right because of magnetic fluctuation or noise in the Feshbach detuning.
\subsubsection{shifts due to phase noise}
When Feshbach coupling strength, $\tilde{g}$ corrupted by Gaussian white noise, $n_x$ then mean, $\langle n_x\rangle=0$, and variances, $\sigma^2_x=\langle n^2_x\rangle=\gamma_x/2$, which generate
\begin{subequations}
\begin{equation}
  \langle (\tilde{g}+n_x)^2\rangle=g^2+\frac{\gamma_x}{2}  
\end{equation}
\text{From Eq. (\ref{shift}), we obtain}
\begin{equation}
    B_{\text{res}}-B_0=\Delta B \bigg(1+\alpha(\tilde{g}^2+\frac{\gamma_x}{2})\bigg) 
\end{equation}    
\end{subequations}
As a result of phase noise, the resonance point shifts to the right side.
\section{Conclusion}\label{conclusion}

In this article, we have studied the noise-induced dynamics of weakly Feshbach-coupled atomic–molecular bosons in a Bloch-sphere framework, with an emphasis on the effects of population imbalance and detuning. The dynamics is analyzed using two complementary approaches:  the MF framework, and the more detailed BBR method that incorporates not just the means but also the second-order correlations of the Bloch vectors' components. 

In our study, the polarization and detuning are treated as the adjustable variables: parameters that can be minutely controlled and measured in present-day experiments. The relaxation behavior of the system is characterized in terms of the longitudinal relaxation time ($T_z$) associated with the decay of polarization ($s_z$), and the transverse relaxation time ($T_{x\text{-}y}$) associated with the decay of coherence ($s_\parallel$). We find that $T_z$ decreases if the system departs more from its equilibrium polarization: something that can be characterized using the drift speed as a marker; and reaches a maximum near Feshbach resonance that is also reflected in the conversion efficiency. In contrast, $T_{x\text{-}y}$ decreases with an increasing imbalance, and reaches its maximum near the resonance: behaviors showing their signatures on the Von Neuumann entropy, and Condensate fraction respectively. 

 In the current study, it is observed that the overall trends with respect to polarization and detuning remain unchanged while going from MF to BBR. This qualitative agreement stems from the fact that the noise in our model is an additive one, well-equipped to explain processes like magnetic/laser field fluctuations. BBR corrections here thus modify amplitudes and timescales only, but they do not affect the qualitative dynamics.

In one of our previous works, the influence of fluctuations and correlations was studied for the same system \cite{mukherjee2025relaxation}. It was found that  when the population imbalance relaxes to a stable equilibrium, the presence of higher-order correlations yields relaxation rates typically lower than the MF predictions. This effect is caused by the suppression of specific decoherence channels via formation of a structured noise: leading to a more robust non-equilibrium configuration. In the current, study, too: this trait is prominent  especially near the resonance; and also near the equilibrium (in the context of relaxation of the polarization), and near the equatorial plane of the Bloch sphere (in the context of relaxation of coherence). Away from resonance, the dominant decay channels are same for both BBR and MF, and the relaxation time nearly matches.

The current study should open up the way for more theoretical and experimental works. There are already a few experimental investigations about relaxation in generic ultracold systems. For example, the relaxation dynamics of Josephson oscillations were observed in a double-well potential using $^{87}\mathrm{Rb}$ atoms (\cite{BJJ1}), where the relative phase and population imbalance were found to relax to a phase-locked state, largely independent of the initial conditions and system parameters \cite{zhang2021transition}. The magnetic field dependence of atom-dimer relaxation and three-body recombination rates were studied in \cite{smirne2007collisional}. Given that the realization of a resonant system of coupled  A-BEC and M-BEC is highly achievable \cite{papp2006observation,zhang2021transition}, it is certain that the relaxation processes in such systems can become extremely important from the experimental perspective. The Hamiltonian in the atom–dimer system has a more interesting nature of nonlinearity compared to their double-well counterparts, and is thus destined to offer a more intricate physics. For future experiments aimed at this direction, this particular work can be useful in identifying the parameter regimes and the relevant time-scale in which the dynamics would be interesting.

\setcounter{equation}{0}
\appendix
\section{Signature of closed system}{\label{closed}}
\begin{figure}[h]
\centering
    \includegraphics[width=0.8\linewidth]{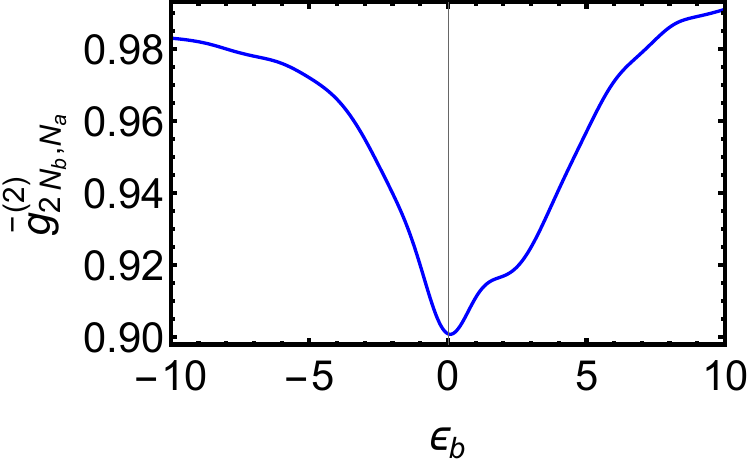}
\caption[short description]{Time-averaged 2nd-order correlations ($g^{(2)}_{2N_b,N_a}$
 ) vs. detuning ($\epsilon_b$), for BBR.}
\label{2nd_correlation}
\end{figure}

The time-averaged second-order correlation function between atom-dimer states is defined as \cite{similar_hamiltonian2,kordas2015dissipative}
\begin{equation} \label{classical correlations} \bar{g}^{(2)}_{2N_b,N_a} = \frac{\overline{2N_b N_a}}{ 2\bar{N}_b \bar{N}_a} = 1 - \frac{\bar{\Delta}_{zz}}{2(1-\bar{s}_z^2)} < 1. \end{equation}
Here, $\bar{g}^{(2)}{2N_b,N_a} < 1$ because coherent population transfer occurs, meaning that atoms and dimers are not created simultaneously; in other words, the creation of one species suppresses the formation of the other, as shown in Fig. (\ref{2nd_correlation}). Here, we obtain the minima of this curve near resonance ($\epsilon_b=0$).\\
From expression of $N$, and $L_z$, we obtain $N_a/N=(1-s_z)/2$, and $2N_b/N=(1+s_z)/2$ \cite{motohashi2010particle,saha2023phase}. So, time dependence of $N_b(t)$ looks like
\begin{equation}
    N_b(t)\approx\frac{N}{2}L_z(t)
\end{equation}
where the constant part is not considered in the dynamics. If we model $N_b(t) = N^0_b e^{-t/T_b}$ and $L_z(t) = L^0_z e^{-t/T_z}$, then $T_b = T_z$, where $N^0_b$ and $L^0_z$ are the amplitudes of $N_b$ and $L_z$, respectively. Here, $T_b$ represents the relaxation time of $N_b$. If we start from a fully molecularly polarized state, corresponding to the North Pole of the Bloch sphere, then the atomic and molecular populations evolve as 
\begin{equation}
  N_a(t) = N(1 - e^{-t/T_a}) \quad \text{and}\quad 2N_b(t) = N e^{-t/T_b}  
\end{equation}
Substituting these expressions into $\dot{N}=0$, we find that $T_a = T_b$ since the system is closed. 

Note that in our system ($^{87}\mathrm{Rb}$), the atomic population increases while the molecular population decreases over time \cite{motohashi2010particle}.
\section{Neglecting three-body and higher order losses}\label{dissipation}
 In an experiment with $^{87}\mbox{Rb}$, the rate constants for various $m$-body scattering processes were estimated.  Near resonance, the two-body loss rates for atom–molecule ($\mathcal{L}_{\text{am}}$) and molecule–molecule ($\mathcal{L}_{\text{mm}}$) collisions are $2\times10^{-10}\text{cm}^6/\text{s}$ and $3\times10^{-10}\text{cm}^6/\text{s}$, respectively \cite{syassen2006collisional}.
In contrast, the three-body recombination loss rate for non-condensed and condensed particles were
$\mathcal{L}^{\mathrm{nc}}_{3} = 1.0 \times 10^{-29}\,\mathrm{cm}^6/\mathrm{s}$ and
$\mathcal{L}^{\mathrm{c}}_{3} = 5.8 \times 10^{-30}\mathrm{cm}^6/\mathrm{s}$
respectively \cite{burt1997coherence}. Near resonance, $\mathcal{L}_{3}$ reaches $10^{-25}\text{cm}^6/\text{s}$ \cite{large_N1}.\\
The decreasing differential equation for the number of atoms becomes \cite{ferlaino2009evidence,weber2003three}, 
\begin{equation}
\frac{1}{N}\frac{dN}{dt}=-\mathcal{L}_2N-\mathcal{L}_3N^2-\mathcal{L}_4N^3   
\end{equation}
where, $\mathcal{L}_i$ stands for $i$-th body loss.\\
Per atom loss rate, \begin{equation}
    \mathcal{G}_i=\mathcal{L}_{i}N^{i-1}
\end{equation}
Two-body and three-body loss processes thus occur on characteristic timescales of
$10^{3}\,\mathrm{s}$ and $10^{25}\,\mathrm{s}$, respectively.

The energy scales used in this article leads to a time scale of the order of $\mu$s. In our case, the dynamics evolves 
for at most $10$ time units.
Within this temporal window, both two-body and three-body loss processes can be safely neglected. Since four-body and higher order loss factors are significantly smaller than three-body loss, these are ignored as well. However, since the dynamics considered here occur within  10 $\mu$s, these losses can be neglected.
\section{Preparation of Fock state}\label{molecule}
\begin{figure}
\centering
\begin{subfigure}{0.8\linewidth}
    \centering
    \includegraphics[width=\linewidth]{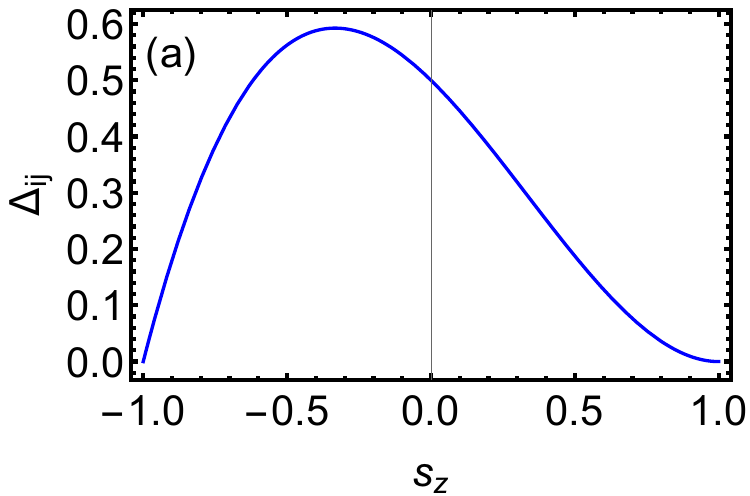}
    \phantomcaption 
    \label{variance_imbalance}
\end{subfigure}
\begin{subfigure}{0.8\linewidth}
    \centering
    \includegraphics[width=\linewidth]{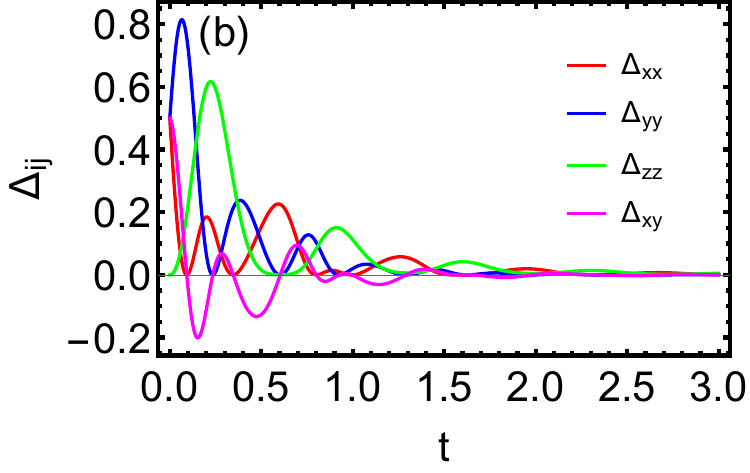}
    \phantomcaption 
    \label{correlation_dynamics}
\end{subfigure}
\caption[short description]{(a) covariance ($\Delta_{ij}$) vs imbalance ($s_z$), and (b) dynamics of $\Delta_{ij}$.}
\label{molecular density}
\end{figure}

If we consider the atomic-molecular Fock state with $N_b$ molecules, $\ket{N-2N_b,N_b}$ \cite{khripkov2011quantum}, then the expectation values of all three components of the Bloch vector are,
\begin{subequations}
\begin{equation}
 \braket{\hat{L}_x}=\braket{\hat{L}_y}=0, \quad\text{and}\quad \braket{\hat{L}_z}=\frac{4N_b}{N}-1   
\end{equation},
\text{and its variances and covariance are considered in} {the large N limit. \cite{strecker2003conversion}},
\begin{equation}
\label{dynamics correlation}
\Delta_{xx}=\Delta_{yy}=\Delta_{xy}=\frac{8N_b}{N}(1-\frac{2N_b}{N})^2,\quad \text{and}\quad \Delta_{zz}=0    
\end{equation}    
\end{subequations}
We plot the covariances $\Delta_{ij}$ versus the imbalance $s_z$ in Fig. \ref{variance_imbalance}, with the variances denoted by $\Delta_{ii}$ for $i=j$. For ${}^{87}\mathrm{Rb}$, the atomic state is energetically favoured over the molecular state \cite{motohashi2010particle}. Consequently, during the evolution, the molecular population 
$2N_b/N$ decreases and eventually vanishes. As a result, the fluctuations $\Delta_{ii}(t)$ approach zero 
in the presence of noise (see Fig. \ref{correlation_dynamics}). This is also evident from Eq. (\ref{dynamics correlation}): as $2N_b/N\to0$, the 
correlations satisfy $\Delta_{xx}\to0$, $\Delta_{yy}\to0$ and $\Delta_{xy}\to0$. Hence, non-Gaussian correlations do not persist in this regime, and Wick's theorem remains valid.
\section{ Expressions of frequency, damping rate and forces}\label{appendixA}
From, Eq. (\ref{EOM}), we obtain the damping rate of Bloch vector components are:\\
\begin{subequations}
    \begin{equation}
  \lambda_x = \frac{\gamma_z}{2}
\end{equation}
    \begin{equation}
 \lambda_y =2\gamma_x(1-3s_z)+\frac{\gamma_z}{2}
    \end{equation}
    \begin{equation}
        \lambda_z =2\gamma_x(1-3s_z)
    \end{equation}
    \label{decay rate}
\end{subequations}
The explicit forms of the squared frequencies are provided in Eq. (\ref{EOM}).\\
\begin{subequations}
    \begin{equation}
\omega^2_x=2c^2_1\Delta_{zz}-(2c_1s_z+c_2)^2 
    \end{equation}
    \begin{equation}
    \begin{split}
\omega^2_y=&2(c^2_1-9\gamma^2_x)-9\gamma^2_x\}\Delta_{zz}+(c_2+2c_1s_z)^2\\&-2\sqrt{2}\tilde{g}(6\gamma_x s_y+3\sqrt{2}\tilde{g}s_z-2c_1s_x-\sqrt{2}\tilde{g})\\&+6\gamma^2_x(1+3s_z)(1-s_z)
    \end{split}
    \end{equation}
    \begin{equation}
\omega^2_z=2s_z\bigg(2\sqrt{2}\tilde{g}c_1s_x+\tilde{g}^2(2-3s_z)-3\gamma_x(s_y+3\gamma_x \Delta_{zz})\bigg) 
    \end{equation}
    \label{omega}
\end{subequations}
From Eq. (\ref{EOM}) we obtain the intrinsic forces on Bloch vector components are:\\
\begin{subequations}
    \begin{equation}
    \label{forcex}
    \begin{split}
     F_x=&\Delta_{zz}\bigg(9\gamma_x c_1 s_y +\frac{3\sqrt{2}\tilde{g}}{2}(3s_zc_1+\frac{c_2}{2})-\sqrt{2}\tilde{g}\bigg)\\&-2c_1(2c_1s_z+c_2)\Delta_{zx}+2\sqrt{2}\tilde{g}c_1\Delta_{yy}\\&+\Delta_{yz}\Bigg(3\gamma_x c_2-\bigg(\frac{\gamma_z}{2}+2\gamma_xc_1(2+9s_z)\bigg)\bigg)\\&+(2c_1s_z+c_2)\bigg(2\gamma_xs_y(3s_z-1)-\frac{\gamma_z s_y}{2}\\&-\frac{\tilde{g}}{\sqrt{2}}(1+2s_z-3s^2_z)\bigg)\\&+2c_1s_y\bigg(2\sqrt{2}\tilde{g}s_y-\gamma_x(1+2s_z-3s^2_z)\bigg)   
    \end{split}
       \end{equation}
       \begin{equation}
       \label{forcey}
           \begin{split}
F_y=&\Delta_{yz}\Bigg(2\bigg(3\tilde{g}^2-c_1(2c_1s_z+c_2)\bigg)-3\gamma_x\bigg(\frac{\gamma_z}{2}+4\gamma_x(1-3s_z)\bigg)\Bigg)\\&-2\sqrt{2}\tilde{g}c_1\Delta_{xy}+\Delta_{zx}\Bigg(c_1\bigg(2\gamma_x(1-3s_z)+\frac{\gamma_z}{2}\bigg)-3\gamma_x(2c_1s_z+c_2)\Bigg)\\&-3\gamma_x\Delta_{zz}\bigg(\sqrt{2}\tilde{g}(1-3s_z)(\gamma_x+\frac{3}{2})+c_1s_x(1+2\gamma_x)\bigg)\\&+6\sqrt{2}\tilde{g}\gamma_x\Delta_{yy}+\frac{\gamma_z s_x}{2}(2c_1s_z+c_2)\\&+\gamma_x(1+s_z)(1-3s_z)\bigg(\sqrt{2}\tilde{g}(1-3s_z)+2c_1s_x\bigg)              
           \end{split}
       \end{equation}
       \begin{equation}
       \label{forcez}
    \begin{split}
F_z=&3\bigg(4\sqrt{2}\tilde{g}\gamma_x\Delta_{yz}+(\tilde{g}^2-2\gamma^2_x)\Delta_{zz}\bigg)\\&-2\tilde{g}\Bigg(\sqrt{2}c_1\Delta_{zx}+\sqrt{2}\bigg(c_2s_x+s_y(2\gamma_x+\frac{\gamma_z}{2})\bigg)+\tilde{g}\Bigg)       
    \end{split}       
       \end{equation}
\end{subequations}
\begin{table*}[t]
\centering
\captionsetup{justification=centering}
\caption{Physical parameters for the system \cite{BEC9,kohler2006production,volz2003characterization}}
\begin{tabular}{|c|c|c|c|c|c|c|c|c|c|c|c|c|c|c|c|}
\hline
$m_{\text{a}}$ (kg) & $\omega$ (Hz) & $L_0$ (m) & $a_{\text{bg}}$ ($a_0$) & $a_{aa}$ ($a_0$) & $a_{bb}$ ($a_0$) & $a_{ab}$ ($a_0$) & $\mu_{\text{co}}$ $(\mu_B)$ & $B-B_0$ (G) & $B_0$ (G) & $\Delta B$ (G)&  $N_a$ (N) & $N_b$ (N)& $\tilde{T}$ $(\mu K)$ \\
\hline
$1.4\times10^{-25}$ & 100 & $10^{-6}$ & $100$ & $58$ & $10^3$ & $-180 \pm 150$ & 2 & 0.5 & 1007.4& 0.21 &2/3 & 1/3 & 10 \\
\hline
\end{tabular}
\label{1st table}
\end{table*}

\begin{table*}[t]
\centering
\captionsetup{justification=centering}
\caption{Bare interaction parameters for the system \cite{BEC9,li2010nonlinear}}
\begin{tabular}{|c|c|c|c|c|c|}
\hline
Quantity & $u_1\,(\text{J}\,\text{m}^3)$ & 
$u_2\,(\text{J}\,\text{m}^3)$ & 
$u_3\,(\text{J}\,\text{m}^3)$ & 
$g\,(\text{J}\,\text{m}^{3/2})$ & 
$\epsilon_b\,(\text{J})$  \\
\hline
Expression & $4\pi\hbar^2 a_{\text{aa}}/m_{\text{a}}$ & 
$4\pi\hbar^2 a_{\text{bb}}/m_{\text{b}}$ & 
$4\pi\hbar^2 a_{\text{ab}}/m_{\text{ab}}$ & 
$\sqrt{u_1 \,\Delta B \,\mu_{\text{co}}}$ & 
$\mu_{\text{co}}(B - B_0)$ \\
\hline
Value & $5.1\times10^{-51}$ & $2.6\times10^{-50}$ & $-5.1\times10^{-51}$ & $2.3\times10^{-30}$ & $9.27\times10^{-28}$\\
\hline
\end{tabular}
\label{2nd table}
\end{table*}
\begin{table}[h]
\centering
\captionsetup{justification=centering}
\caption{Dimensionless effective parameters}
\begin{tabular}{|c|c|c|c|}
\hline
Parameter & Expression& Value (J) & Scaling \\
\hline
$U_1$ & $u_1 N_a / V$ & 1.7$\times10^{-28}$ &1 \\
\hline
$U_2$ & $u_2 N_b / V$ & $7.3\times10^{-28}$ & 4.3 \\
\hline
$U_3$ & $u_3 N / V$& $1.19\times10^{-27}$ & -7 \\
\hline
$\tilde{g}$ & $g\sqrt{N_a/V}$ &  3.4$\times10^{-28}$ &2 \\
\hline
$\epsilon_b$ & $\mu_{\text{co}}(B-B_0)$ & $1.53\times10^{-27}$ & 9.1 \\
\hline
\end{tabular}
\label{3rd table}
\end{table}
\begin{table}[h]
\centering
\captionsetup{justification=centering}
\caption{Thermal parameters of the system \cite{dalvit2000decoherence,liu2010shapiro}}
\begin{tabular}{|c|c|c|c|c|c|}
\hline
$N_{\text{th}}$ & $n_{\text{th}}$ ($\mathrm{m}^{-3}$) & $v_{\text{th}}$ (m$\text{s}^{-1}$) & $\Gamma_x$ (KHz) & $\tilde{\Gamma}_x$ (MHz) & $\gamma_x$ (J) \\
\hline
$10^3$ & $5\times10^{19}$ & $1.4\times10^{-2}$ & $0.5$ & $0.1$ & $1.3\times10^{-29}$ \\
\hline
\end{tabular}
\label{4th table}
\end{table}

\section{Choice of initial state, and experimental parameters} \label{initial state}
 The scattering lengths $a_{\text{aa}}$ and $a_{\text{bb}}$ correspond to atom–atom and molecule–molecule interactions, respectively, while $a_{\text{ab}}$ denotes the atom–molecule scattering length; all are expressed in units of the Bohr radius ($a_0$) in Table \ref{1st table}. The reduced mass of the atom–molecule pair is given by $m_{\text{ab}} = m_a m_b/(m_a + m_b)$, where $m_a$ and $m_b$ are the atomic and molecular masses, respectively. The characteristic length scale of the harmonic trap is \cite{BEC9} $L_0 = \sqrt{\hbar/m_{a}\omega}$, where $\omega$ is the trap frequency. From Appendix \ref{closed}, $N_a$ and $N_b$ can be written in terms of $N$ and $s_z$. Upon setting $s_z = 0$, they depend only on $N$. All fundamental parameter values are listed in Table \ref{1st table}. The trap volume is therefore estimated as $V \sim L_0^3 = 2 \times 10^{-17}\mathrm{m}^3$. The bare and effective interaction strengths are listed in Tables \ref{2nd table} and \ref{3rd table}, respectively.\\
The collision rate ($\Gamma_x$) between condensed and thermal atoms is given by $8\pi a_{\text{eff}}^{2} n_{\text{th}} v_{\text{th}}$ \cite{liu2010shapiro}, where $v_{\text{th}}$ is the thermal velocity associated with the temperature $\tilde{T}$. The effective scattering length is $a_{\text{eff}}$ yielding the identical-boson cross section $8\pi a_{\text{eff}}^{2}$ \cite{liu2010shapiro}. The corresponding diffusion rate is $\tilde{\Gamma}_x = 8\pi^3 (8\pi a_{\text{eff}}^{2} n_{\text{th}} v_{\text{th}})$, where $n_{\text{th}}$ is the thermal density corresponding to $N_{\text{th}}$ per unit quantization volume \cite{liu2010shapiro}. Using these parameters, the relaxation energy scale is $\gamma_x=\hbar\tilde{\Gamma}_x \approx 0.1\tilde{g}$ (see Table \ref{4th table}). We further assume magnetic-field fluctuations of $10\%$ of $B$, yielding $\gamma_z = 0.1\epsilon_b$, where $\mu_{\text{co}}$ is measured in units of the Bohr magneton ($\mu_B$).

\section{Characteristic Polynomial Framework for System Stability}\label{appendix B}
 \text{If $\Delta_{ij}=0$, then Eq. (\ref{covariance}) generates the Jacobian matrix ($J_{ij}$),}
 \begin{equation}
 \begin{pmatrix}
-\frac{\gamma_z}{2} & 2c_1s_z+c_2 & 2c_1s_y\\
-(2c_1s_z+c_2) & 2\gamma_x(3s_z-1)-\frac{\gamma_z}{2} & -2c_1s_x-\sqrt{2}\tilde{g}(1-3s_z)\\
0 & 2\sqrt{2}\tilde{g} & 2\gamma_x(3s_z-1)
 \end{pmatrix}    
 \end{equation}
 Using the two equilibrium points, unstable $(0,0,1)$ and stable $(0,0,-1/3)$, we obtain the characteristic equations are:
 \begin{subequations}
 \label{polynomial}
\begin{equation}
\begin{split}
 &(\lambda+\frac{\gamma_z}{2})\bigg((\lambda-4\gamma_x+\frac{\gamma_z}{2})(\lambda-4\gamma_x)-8\tilde{g}^2\bigg)\\&+k_1^2(\lambda-4\gamma_x)=0
 \end{split}
\end{equation}
\text{where, $k_1=c_2-2c_1$}
\begin{equation}
\begin{split}
&(\lambda+\frac{\gamma_z}{2})\bigg((\lambda+4\gamma_x+\frac{\gamma_z}{2})(\lambda+4\gamma_x)+8\tilde{g}^2\bigg)\\&+k_2^2(\lambda+4\gamma_x)=0.      
\end{split}    
\end{equation}
\text{where, $k_2=c_2-2c_1/3$}
 \end{subequations}

\bibliography{bibi.bib}

\end{document}